\documentclass[superscriptaddress, twocolumn, prd]{revtex4-1}
\usepackage[latin9]{inputenc}
\usepackage{amsmath}
\usepackage{amstext}
\usepackage{graphicx}
\usepackage{physics}
\usepackage{changes}

\begin{document}

\title{A Pseudo Plane-wave Gravitational Calibrator for Gravitational Wave Observatories}

\author{M.P. Ross}
\email[]{mpross2@uw.edu}
\author{J.H. Gundlach}
\email[]{gundlach@uw.edu}
\author{E.G. Adelberger}
\affiliation{Center for Experimental Nuclear Physics and Astrophysics, University of Washington, Seattle, Washington,
98195, USA}
\author{C.M. Weller}
\affiliation{California Institute of Technology, Pasadena, CA, 91125, USA}
\author{E.A. Shaw}
\author{C. Gettings}
\affiliation{Center for Experimental Nuclear Physics and Astrophysics, University of Washington, Seattle, Washington,
98195, USA}
\author{J. Kissel}
\affiliation{LIGO Hanford Observatory, Richland, WA 99352, USA}
\author{T. Mistry}
\affiliation{The University of Sheffield, Sheffield S10 2TN, UK}
\author{L. Datrier}
\affiliation{SUPA, University of Glasgow, Glasgow G12 8QQ, UK}
\author{E. Daw}
\affiliation{The University of Sheffield, Sheffield S10 2TN, UK}
\author{M. Hendry}
\affiliation{SUPA, University of Glasgow, Glasgow G12 8QQ, UK}

\begin{abstract}

 The precisions of existing gravitational calibrators for gravitational wave observatories are limited by their dependence on the relative position between the calibrators and the observatory's test masses. Here we present a novel geometry consisting of four quadrupole rotors placed at the vertices of a rectangle centered on the test mass. The phases and rotation directions are selected to produce a pseudo plane-wave  sinusoidal gravitational acceleration with amplitude of $\sim100$ fm/s$^2$. We show that this acceleration only has minimal dependence on the test mass position relative to the rotor array and can yield $0.15 \%$ acceleration amplitude uncertainty while tolerating a 1-cm test mass position uncertainty. The acceleration can be directed precisely along the optical axis of the interferometer arm and applies no torque on the test mass. In addition, the small size of the rotors has significant engineering and safety benefits.

\end{abstract}

\maketitle

\section{Introduction}

Gravitational wave astronomy has blossomed into a novel method to observe the universe. The number of gravitational wave observations is expected to grow substantially in the coming years with the continued operation of the LIGO \cite{aLIGO} and Virgo \cite{virgo} interferometers as well the future addition of LIGO-India \cite{ligo-india} and the further improvements of KAGRA \cite{kagra}.

Precise and robust absolute calibration of these interferometers is essential. Cosmological measurements \cite{abbott2021gravitational, ligo2017gravitational, schutz1986determining}, searches for deviations from general relativity \cite{abbott2020tests}, and binary-merger characterization \cite{abbott2020population} all require precise strain calibrations. Currently these calibrations are made using photon pressure \cite{PCal}. These calibration systems provide absolute calibrations limited to $\sim0.4\%$ uncertainty \cite{Bhattacharjee_2020}. Improvements on this have proven difficult. In addition, relying on a single calibration system may be susceptible to unknown systematics.

%These interferometers must be precisely calibrated to accurately interpret gravitational wave signals. Whether for binary merger parameter estimation \cite{abbott2020population}, cosmological measurements \cite{abbott2021gravitational, ligo2017gravitational, schutz1986determining}, or searches for deviations from general relativity \cite{abbott2020tests}, the strain readouts of the observatories must be precisely and accurately calibrated to yield accurate scientific results.
%
%Currently, this calibration has been accomplished using photon pressure \cite{PCal}. These versatile photon calibration systems have yielded absolute calibrations with $\sim0.41\%$ uncertainty \cite{Bhattacharjee_2020}. However, relying on a single calibration system can allow unknown systematics to skew the scientific results.

Calibrating with a gravitationally induced strain has long been suggested as an alternative calibration technique \cite{hirakawa1980dynamical, kuroda1985experimental, mio1987experimental, astone1991evaluation, astone1998experimental, Matone_2007} and has recently been implemented at gravitational wave observatories \cite{Estevez_2018, estevez2021newtonian, PhysRevD.98.022005, ncal}. Gravitational calibration has multiple advantages over photon pressure including minimal sources of systematic error and guaranteed stability over long time durations. Operating both gravitational and photon calibration systems allows the systems to cross-check each other and combined yield a higher-precision absolute calibration.

Single-rotor gravitational calibrators \cite{Estevez_2018, PhysRevD.98.022005, ncal} produce accelerations that have large dependence on the radial distance, $r$, between the rotor and the test mass. The acceleration is typically proportional to $\sim1/r^{l+2}$ where $l$ is the order of the dominant mass-multipole moment. For example, a rotor with a quadrupole mass distribution ($l=2$) will follow $\sim1/r^4$. This strong positioning dependence causes the performance of the absolute calibration to be limited by the measurement of the test mass to rotor separation. The Virgo observatory is pursuing an asymmetric system of rotors rotating at different frequencies that may alleviate this limitation \cite{estevez2021newtonian}.

\begin{figure}[!h]
\centering \includegraphics[width=0.45\textwidth]{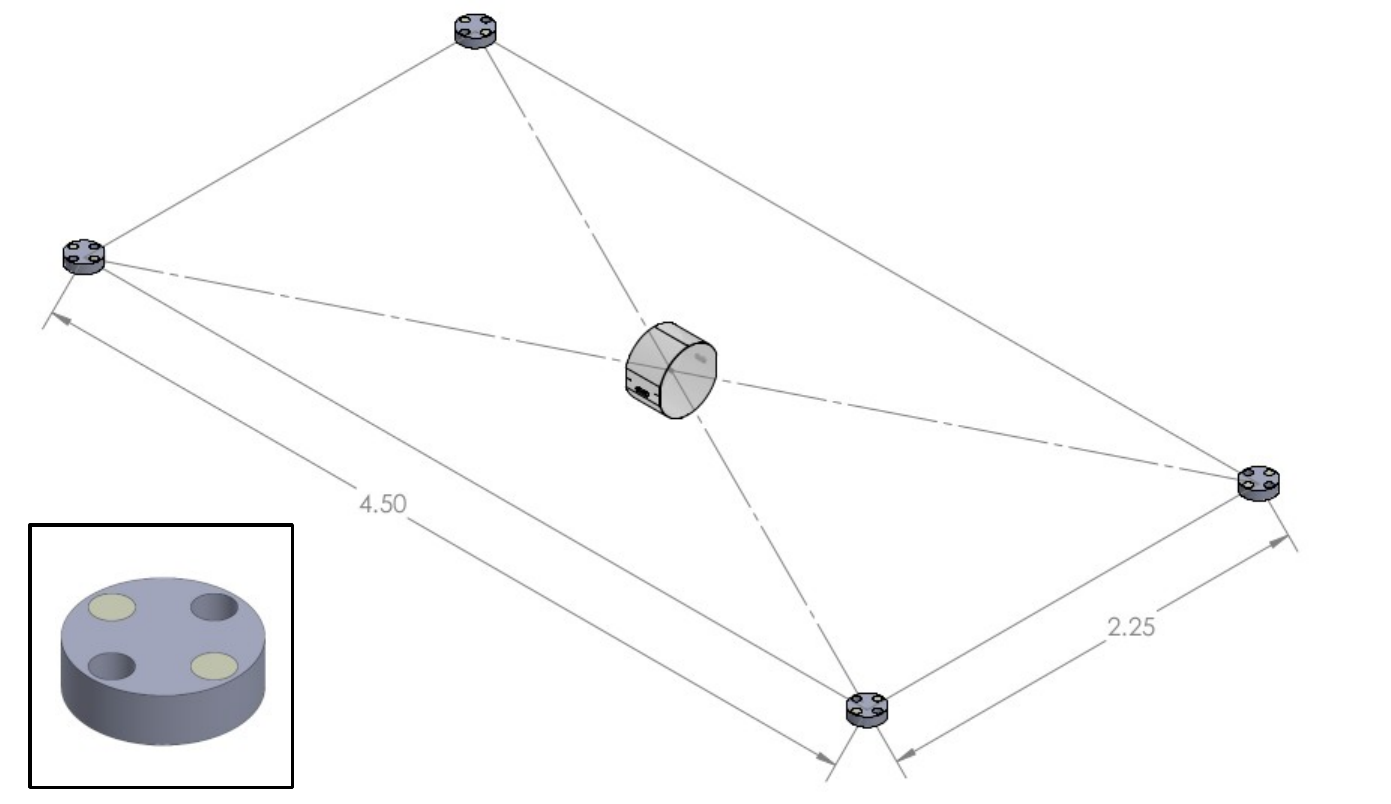}
\caption{A rendering of the geometry of the rotors with the test mass at the center of the 2.25-m by 4.50-m rectangle. Also shown is a detail of a single rotor consisting of the dark gray aluminum disk with two light gray tungsten slugs inserted into two of the four rotor holes.}
\label{cad} 
\end{figure}

Here, we present a novel geometry consisting of four quadrupole rotors that produces a pseudo plane-wave gravitational acceleration, see Section \ref{pseudo}. This symmetric geometry has no first-order dependence on the position of the rotors relative to the test mass. It instead depends on the easy-to-measure positions between the rotors in the array. Additionally, this geometry suppresses the torques acting on the test mass and eases much of the engineering and safety concerns of previous rotors.

%\begin{widetext}
%\begingroup
%\setlength{\tabcolsep}{10pt} % Default value: 6pt
%\renewcommand{\arraystretch}{1.5} % Default value: 1
%
%\begin{table}[h!]
%\begin{center}
%\begin{tabular}{ |l|c|c| }
%\hline
% Parameter & Mean & Uncertainty \\
% \hline
%Cylinder Mass & 1~kg & 0.3~g \\
%Cylinder Radius & 2~cm & 2.5 $\mu$m \\
%Cylinder Length & 5~cm & 5 $\mu$m \\
%Quadrupole Radius & 6~cm & 5 $\mu$m \\
%Test Mass & 40~kg & 10~g \\
%Test Mass Length & 200~mm & 0.1~mm\\
%Test Mass Radius & 170~mm & 0.05~mm\\
%Test Mass Flat Width & 327~mm & 0.05~mm\\
%Rotor Positions & ($\pm$ 2.25 m, $\pm$ 1.125 m, 0 m) & (1 mm, 1 mm, 1 mm) \\
%Test Mass Position & (0 m, 0 m, 0 m) & (1 mm, 1 mm, 1 mm) \\
%Rotor Relative Phase & $0^\circ$, $90^\circ$ & $1^\circ$ \\
% \hline
%
% \end{tabular}
% \caption{Parameters describing the rotors, the test mass, and their respective positions.}\label{param}
% \end{center}
%
%\end{table}
%\endgroup
%\end{widetext}

\begin{figure}[!h]
\centering \includegraphics[width=0.5\textwidth]{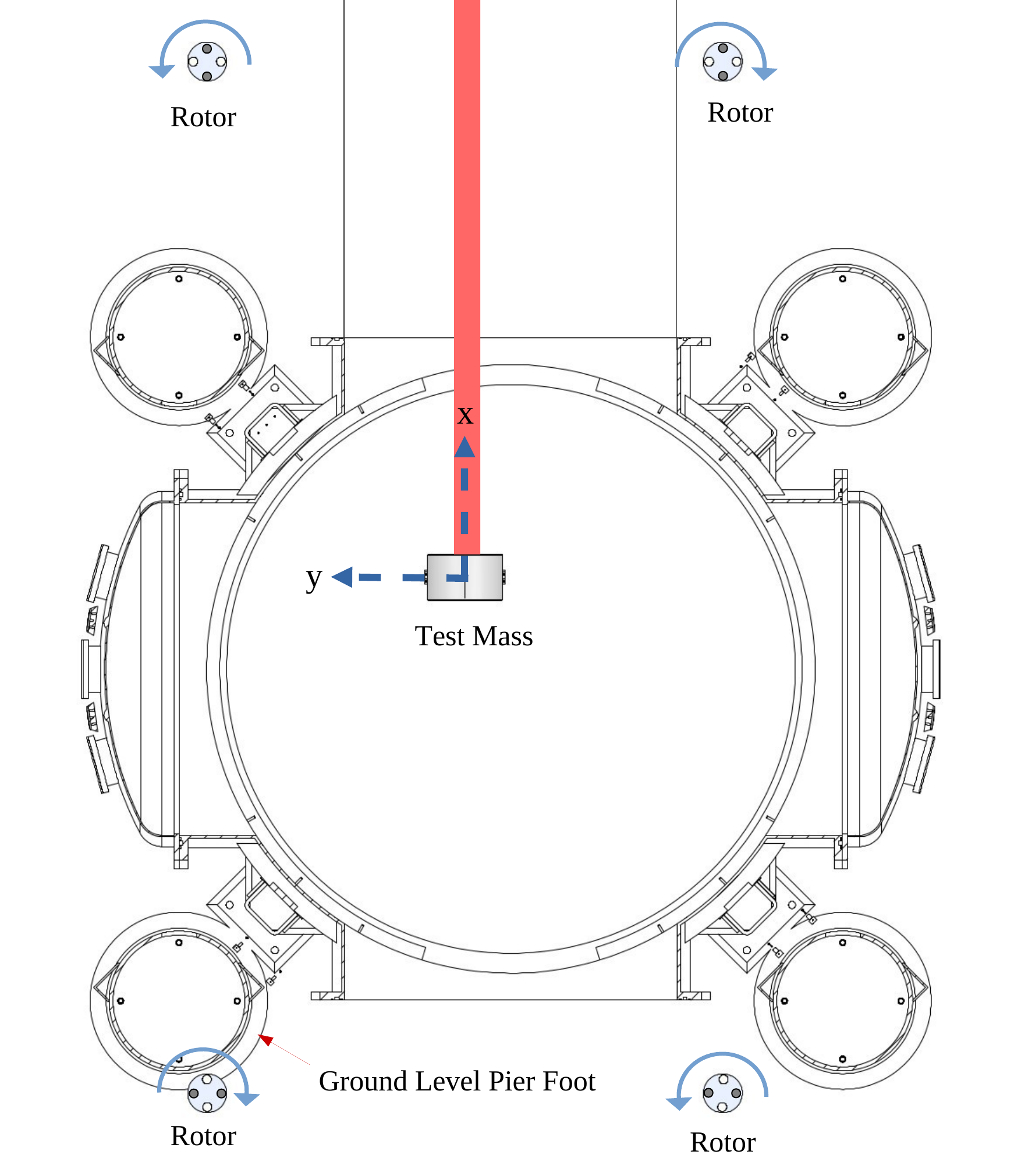}
\caption{A rendering of the geometry of the rotor array around the LIGO BSC chamber with the corresponding test mass at the center of our coordinate system and the observatory's main interferometer beam schematically shown in red. The overlapping ground level pier foot is far below the plane of the rotors and test mass.}
\label{bsc} 
\end{figure}

\section{Geometry}

The example pseudo plane-wave calibrator proposed here consists of four identical rotors placed at the vertices of a 2.25-m by 4.50-m rectangle centered on the test mass. Figure \ref{cad} shows a rendering of the geometry placed around LIGO's end-station vacuum chamber. The rotors are designed with the similar dimensions as the LIGO NCal \cite{ncal} but without a hexapole mass arrangement. Each rotor is a 17-cm diameter, 5-cm tall aluminum disk with four holes cut at a radius of 6-cm separated by $90^\circ$. Two holes are filled in with 4-cm diameter, 5-cm tall  tungsten cylinders which produce a quadrupole mass distribution. The parameters of the geometry are displayed in Table~\ref{uncert}.

The rotor parameters that are common with the LIGO NCal are assigned uncertainties equal to what was previously achieved \cite{ncal}. The rest of the parameters (positioning, phase, etc.) are assigned uncertainties based on what is reasonably achievable with standard measurement techniques. For example, since the rotors would be outside the interferometer's vacuum system, their positions can be readily measured to  mm-precision with standard surveying equipment \cite{ncal}.

The relative phases of the rotors and the rotation directions are set to achieve a pseudo plane-wave nature. The rotors with positive x-coordinates are rotated by $90^\circ$ relative to the rotors with negative x-coordinates. Additionally, the rotors with matching x- and y-coordinate signs (i.e. both positive or both negative) rotate counter-clockwise while those with mismatched signs rotate clockwise. These choices cause the applied acceleration to be purely in the x-direction with no net torques.

\section{Engineering Simplicity}

Since the four rotor array produces more acceleration at a given separation than a single rotor, the array can be placed at a larger radius to produce a similar amplitude acceleration on the test mass. This allows the array to be placed well away from the existing infrastructure of the observatories. Here we have chosen a geometry that fits around the LIGO BSC vacuum chamber and seismic isolation system, as shown in Figure \ref{bsc}. This significantly simplifies the structure that holds the rotors as it does not need to be incorporated into the existing structural components. 

The use of only quadrupole mass distributions as compared to both quadrupole and hexapole masses \cite{ncal} decreases the rotor's kinetic energy thus decreasing the likelihood of damage through catastrophic failure. Due to the decreased moment of inertia, a smaller radius also decreases the torques needed to spin the rotor and maintain a fixed rotation speed which loosens the requirements on the drive motors as well as decreases spurious electromagnetic effects that caused by the motors and auxiliary electronics.

%Cyl Mass: 101.4239 fm/s^2 +- 0.035909 fm/s^2 (0.035405 % )
%Cyl Radius: 101.4229 fm/s^2 +- 1.5286e-06 fm/s^2 (1.5071e-06 % )
%Cyl Length: 101.4229 fm/s^2 +- 2.7807e-06 fm/s^2 (2.7417e-06 % )
%Quad Radius: 101.4202 fm/s^2 +- 0.017544 fm/s^2 (0.017298 % )
%TM Mass: 101.4229 fm/s^2 +- 1.8933e-13 fm/s^2 (1.8667e-13 % )
%TM Radius: 101.4228 fm/s^2 +- 0.00039918 fm/s^2 (0.00039358 % )
%TM Length: 101.4229 fm/s^2 +- 0.00040567 fm/s^2 (0.00039998 % )
%TM Width: 101.4229 fm/s^2 +- 1.3884e-13 fm/s^2 (1.3689e-13 % )
%Rotor Position: 101.4294 fm/s^2 +- 0.10658 fm/s^2 (0.10508 % )
%TM Position: 101.4222 fm/s^2 +- 0.013569 fm/s^2 (0.013379 % )
%Rotor Phase: 101.3465 fm/s^2 +- 0.12589 fm/s^2 (0.12422 % )

%Acceleration: 101.367 fm/s^2 +- 0.1548 fm/s^2 (0.15271 % )

\begin{figure}[!h]
\centering \includegraphics[width=0.5\textwidth]{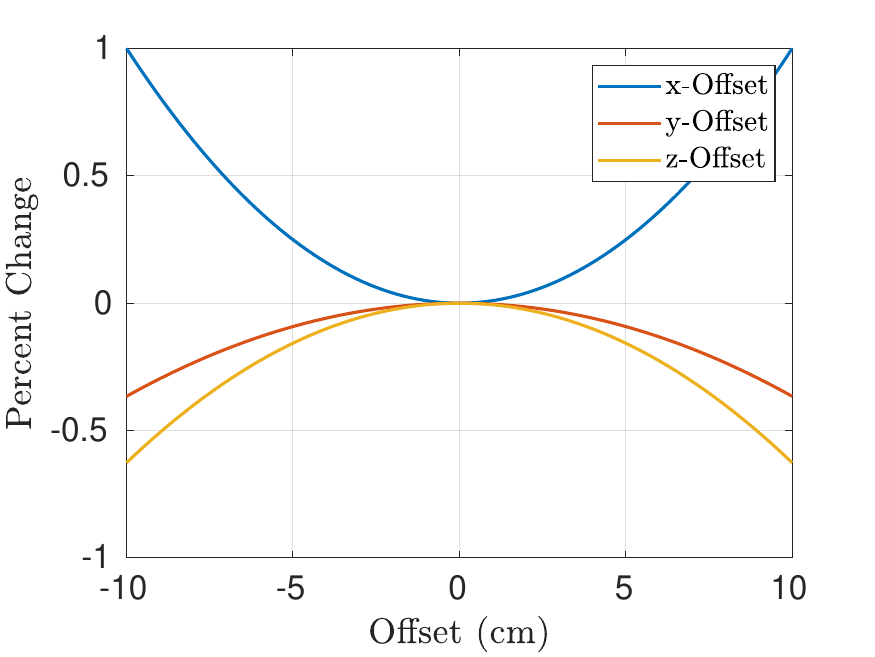}
\caption{The percentage change of the acceleration amplitude applied by the rotor array with a test mass offset from the center of the rectangle. The relatively large offset of 1-cm in any direction causes a $<$1\% change in the acceleration amplitude.}
\label{offset} 
\end{figure}

\begin{widetext}
\begingroup
\setlength{\tabcolsep}{10pt} % Default value: 6pt
\renewcommand{\arraystretch}{1.5} % Default value: 1

\begin{table}[h!]
\begin{center}
\begin{tabular}{ |l|c|c|c| }
\hline
 Parameter & Mean & Estimated Uncertainty & Fractional Acceleration Uncertainty\\
 \hline

Cylinder Mass & 1~kg & 0.3~g & $3.5\times10^{-4}$\\
Cylinder Radius & 2~cm & 2.5 $\mu$m & $1.5\times10^{-8}$\\
Cylinder Length & 5~cm & 5 $\mu$m & $2.7\times10^{-8}$\\
Quadrupole Radius & 6~cm & 5 $\mu$m &$1.7\times10^{-4}$\\
Test Mass* & 40~kg & 10~g & $1.9\times10^{-15}$\\
Test Mass Length & 200~mm & 0.1~mm& $4.0\times10^{-6}$\\
Test Mass Radius & 170~mm & 0.05~mm& $3.9\times10^{-6}$\\
Test Mass Flat Width & 327~mm & 0.05~mm&$1.4\times10^{-15}$\\
Rotor Positions & ($\pm$ 2.25 m, $\pm$ 1.125 m, 0 m) & (1 mm, 1 mm, 1 mm) &$1.1\times10^{-3}$\\
Test Mass Position & (0 m, 0 m, 0 m) & (1 cm, 1 cm, 1 cm) &$1.3\times10^{-4}$\\
Rotor Relative Phase & $0^\circ$, $90^\circ$ & $1^\circ$ & $1.2\times10^{-3}$\\
 \hline
 \hline
& &Quadrature Sum  &$1.68\times10^{-3}$\\
\hline
 \end{tabular}
 \caption{Individual contributions to the acceleration uncertainty for the parameters of the simulation. The quadrature sum is only an approximation of the uncertainty. The full uncertainty must take into account non-linearities and degeneracies as is done in Figure \ref{dist}. *Since the gravitational acceleration is independent of the test mass, this entry represents the numerical precision of the simulation.}\label{uncert}
 \end{center}

\end{table}
\endgroup
\end{widetext}

\section{Pseudo Plane-wave Nature}\label{pseudo}

To verify the performance of such a rotor array, we simulated the system with a finite-element analysis using the \textit{PointGravity} algorithms of the \texttt{newt} libraries \cite{Hagedorn, pgURL}. This simulation breaks each of the rotor cylinders and the test mass into independent clouds of point masses. The force between each pair of point masses, one from the rotors and the other from the test mass, is calculated. The simulation sums the forces between all rotor-test-mass point pairs to yield the acceleration in all three directions. We extract only the x-acceleration as this is the sensitive direction of the interferometer. Although not detailed here, the acceleration predictions were cross-checked with the results of an analytical point-mass approximation \cite{ncal} and an independent  numerical integration calculation.

The superposition of the gravitational fields from the four rotors produces an oscillating gravitational acceleration field which at the center of the rectangle is purely in the x-direction and has an amplitude of $101.37$ fm/s$^2$. This amplitude corresponds to a strain of $7.1\times10^{-22}$ at 30 Hz for the 4-km long interferometer. Note that although the acceleration amplitude is frequency independent, the strain amplitude will follow $\sim1/f^2$. 

The acceleration field changes weakly with deviations from the center of the rectangle (i.e. a pseudo plane-wave). The percentage change in acceleration amplitude versus offset from the center of the rectangle is shown in Figure \ref{offset} for offsets in each direction. A relatively large offset of 10 cm in any direction changes the acceleration by $<1\%$. Additionally, the change in amplitude is well-described by a parabola for small offsets displaying the second-order nature of this effect. 

Since the rotor array is in-plane and symmetric about the x-z plane, the rotors apply no net torque on the test mass. If the array is out-of-plane then the test mass would experience a torque about the y-axis. Similarly, if the array is right-left asymmetric it would apply a z-axis torque. Such torques are common in existing gravitational calibrators and can substantially impact the precision of the subsequent calibrations \cite{ncal}. Note that a different selection of relative rotor phases and rotation directions can apply a net torque on the test mass with no net force. This configuration could provide a novel diagnostic tool for evaluating the interferometer's angular sensitivity and beam spot offsets.

\begin{figure}[!h]
\centering \includegraphics[width=0.45\textwidth]{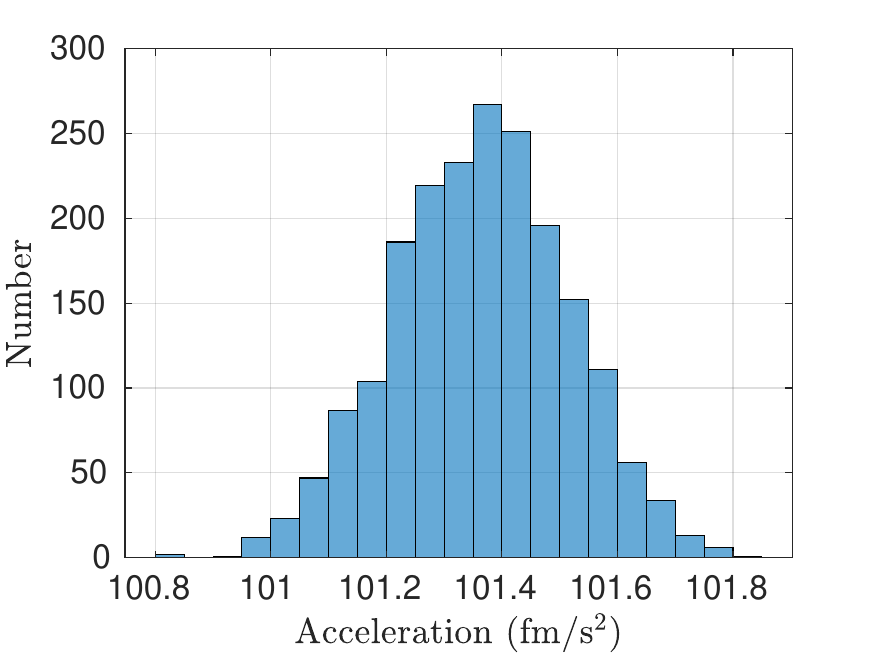}
\caption{Distribution of predicted accelerations for the rotor array described by the parameters in Table \ref{uncert}. This distribution yields an acceleration of 101.37~$\pm$~0.15 fm/s$^2$~(0.15~\%).}
\label{dist} 
\end{figure}
\section{Numerical Uncertainty Analysis}

The ultimate precision of our four-rotor calibrator depends on all the parameters in Table \ref{uncert}. We performed a Monte Carlo simulation of the applied acceleration accounting for the set of parameters which describe the calibrator. We modeled each parameter as a Gaussian distribution centered on the mean listed in Table \ref{uncert} with $\sigma$-value equal to the uncertainty. The acceleration of the test mass was then calculated with parameters sampled from these distributions. This was repeated 2000 times to yield a distribution of the gravitational acceleration, shown in Figure \ref{dist}, which took into account all non-linearities and degeneracies.

The simulation yields an injected acceleration of a~=~101.37~$\pm$~0.15 fm/s$^2$~(0.15~\%) where the central value is the mean and the uncertainty is the 68\%-confidence. To assess how each parameter contributes to this total uncertainty, the acceleration uncertainty was recomputed with only one parameter varying. This was then repeated for each parameter to yield the results in Table \ref{uncert}. All four rotor positions were simultaneously varied  in all three directions and the test mass position was also varied in all three directions.

Table \ref{uncert} shows the acceleration uncertainty is strongly dominated by the rotor positions and relative phases with the test mass position contribution being seven to nine times smaller. These contributions may be further reduced with a higher precision surveying and phase determination than is assumed here.
\\
\section{Conclusion} 
\quad We have described a four-rotor gravitational calibrator that produces a psuedo-plane wave acceleration field, providing a direct and robust absolute calibration with simple systematic uncertainties. Simulation of the acceleration amplitude uncertainty shows that such a system can readily achieve an absolute precision of $0.15\%$. This is approximately an order of magnitude improvement over previously deployed geometries at LIGO \cite{ncal}. In addition to yielding a precise calibration, such a system may be used to search for non-Newtonian gravity \citep{PhysRevD.84.082002}, make terrestrial measurements of Shapiro delay \citep{Ballmer_2010, Sullivan_2020}, and measure the gravitational constant~\citep{NCalGPaper}.
\\

\begin{acknowledgements}

Participation from the University of Washington, Seattle, was supported by funding from the NSF under Awards PHY-1607385, PHY-1607391, PHY-1912380, and PHY-1912514.

\end{acknowledgements}

 \bibliographystyle{unsrtnat}
\bibliography{Super4.bib}

\end{document}